\documentstyle[aps,floats,eqsecnum,psfig]{revtex}

\begin{document}
\draft

\title{A hollow sphere as a detector of gravitational radiation}

\date{\today}
\date{15 March 1997) \\ (Revised \today}

\author{E. Coccia}
\address{Dipartimento di Fisica, Universit\`a di Roma ``Tor Vergata''
	{\rm and\/} \\ INFN Sezione di Roma Tor Vergata, Via Ricerca 
	Scientifica 1, 00133 Roma, Italy}
\author{V. Fafone}
\address{INFN Laboratori Nazionali di Frascati, Via E. Fermi 40, 00044
	Frascati, (Roma), Italy}
\author{G. Frossati}
\address{Kamerlingh Onnes Laboratory, Leiden University, Leiden, The
	Netherlands}
\author{J. A. Lobo {\it and\/} J.A. Ortega}
\address{Departament de F\'\i sica Fonamental, Universitat de Barcelona,
	Spain}

\maketitle

\begin{abstract}

The most important features of the proposed spherical gravitational wave
detectors are closely linked with their symmetry. {\it Hollow\/} spheres
share this property with {\it solid\/} ones, considered in the literature
so far, and constitute an interesting alternative for the realization of
an omnidirectional gravitational wave detector. In this paper we address
the problem of how a hollow elastic sphere interacts with an incoming
gravitational wave and find an analytical solution for its normal mode
spectrum and response, as well as for its energy absorption cross
sections. It appears that this shape can be designed having relatively low 
resonance frequencies ($\sim$\,200 Hz) yet keeping a large cross section, 
so its frequency range overlaps with the projected large interferometers. 
We also apply the obtained results to discuss the performance of a hollow
sphere as a detector for a variety of gravitational wave signals. 

\end{abstract}

\pacs{04.80.Nn, 95.55.Ym}

\section{Introduction}

Thirty-five years after the beginning of the experimental search for cosmic
gravitational waves (GW), several resonant-mass detectors (cryogenic
cylindrical bars) are currently monitoring the strongest potential sources
in our Galaxy and in the local group~\cite{mr}. The sensitivity of such
detectors is $h \simeq$ 6 $\times$ 10$^{-19}$ for millisecond GW bursts, or,
in spectral units,  10$^{-21}$ Hz$^{-1/2}$ over a bandwidth of a few Hz around
1 kHz. A further improvement in sensitivity and bandwidth is expected from the
operation at ultralow temperatures of the two bar detectors {\sl NAUTILUS\/}
\cite{nau} and {\sl AURIGA\/} \cite{aur} in Italy, and even better
sensitivities and bandwidths will come about as more advanced readout systems
are developed. Projects for spherical resonant-mass GW detectors have emerged
in the last few years in the resonant-mass community~\cite{sg,sp,ss,st}, due
to their remarkable advantages with respect to the operating bars~\cite{ll}.

In a cylindrical bar only the first longitudinal mode of vibration interacts
strongly with the GW, and consequently only {\it one\/} wave parameter can
be measured: the amplitude of a combination of the two polarization states.
On the other hand each quadrupole mode of a spherical mass is five-fold
degenerate (its angular dependence is described in terms of the five spherical
harmonics $Y_{lm}$($\theta$,$\varphi$) with $l\/$=2 and $m$\,=\,$-$2,...,2),
and presents an {\it isotropic\/} cross section. The cross section of the
lowest order ($n\/$=1) mode is the highest, and is larger than that of a
cylindrical antenna made of the same material and with the same resonant
frequency by a factor of about $0.8\,(R_{s} / R_{b} )^2$ \cite{ss,st}, where
$R_s$ and $R_b$ are the radius of the sphere and of the bar, respectively.
This means a factor of 20 over present bars. Moreover, the sphere's cross
section is also high at its second quadrupole harmonic.

The five-fold degeneracy of the quadrupole modes enables the determination
of the GW amplitudes of two polarization states and the two angles of the 
source direction. The method first outlined by Forward~\cite{sk} and
later developed by Wagoner and Paik~\cite{sm}, consists in measuring the
sphere vibrations in at least five independent locations on the sphere surface
so as to determine the vibration amplitude of each of the five degenerate
modes. The Fourier components of the GW amplitudes at any quadrupole
frequencies and the two angles defining the source direction can be obtained
as suitable combinations of these five outputs~\cite{sp,ss,ll,tb,ae}.

The signal deconvolution is based on the assumption that in the wave frame
(that in which the $z\/$ axis is aligned with the wave propagation direction)
only the $l\/$=2 and $m\/$=$\pm$2 modes are excited by the GW, as the helicity 
of a GW is 2 in General Relativity. One can take advantage of this to
deconvolve the wave propagation direction and the GW amplitudes in the wave
frame.

Most of the nice properties of a spherical GW detector depend on its being
{\it spherically symmetric\/}. A spherical shell, or {\it hollow sphere\/},
obviously maintains that symmetry, so it can be considered an interesting
alternative to the usual {\it solid\/} sphere. In order to have a good
cross section, a resonant GW detector must be made of a high speed of sound
material, and have a large mass. The actual construction of a massive
spherical body may be technically difficult. In fact, fabricating a large
hollow sphere is a different task than fabricating a solid one. Casting a
hollow half sphere  is a nearly two dimensional cast, at odds with casting
a solid sphere, which requires rather special moulds. As an example of the
feasibility of large two dimensional casting we can mention the fabrication
of propellers of more than 10 meters in size and masses of the order of 100
tons~\cite{lips}. Two hollow hemispheres could then be welded together with
electron beam techniques. However, while it is known that these welding
technique preserve most of the properties of the bare material, its effect
on the acoustic quality factor (a relevant paramenter in resonant mass
detectors) must be further studied.

We have investigated the properties of a hollow sphere as a potential GW
antenna. The purpose of this paper is to present a detailed report of the
main results of such an investigation, and to discuss the real interest of
this new detector shape.

In section 2 we present the complete analytical solution of the eigenmode
problem for a hollow sphere of arbitrary thickness, including the full
frequency and amplitude spectrum. Section 3 is devoted to the cross section
analysis, while in section 4 we take up the study of the system sensitivity
to various GW signal classes. Finally, we present an outlook and summary of
conclusions in section 5.

\section{Normal modes of vibration and eigenfrequencies of a hollow sphere}

In this section we consider the problem of a hollow elastic sphere in order
to obtain its normal modes and frequency spectrum. This is a classical problem
in Elasticity theory which was posed and partly addressed already in the
last century, see e.g. \cite{love} and references therein.

Let $R\/$ and
$a\/$ be the outer and inner radius of the sphere, respectively. The elastic
properties of the sphere, provided it is homogenous and isotropic, will be
described by its Lam\`e coefficients, $\lambda$ and $\mu$, and its density,
$\rho$. As is well known (see, e.g., \cite{ll}), the normal modes are
obtained as the solutions to the eigenvalue equation

\begin{equation}
\nabla^2{\bf u}+\left(1+\lambda/\mu\right)
\nabla (\nabla\cdot{\bf u})=-k^2{\bf u},
\hspace{1cm} \left(k^2\equiv\omega^2\rho/\mu\right),
\label{eigenv}
\end{equation}

subject to the boundary conditions that the solid's surface be free of any
tensions and/or tractions; these are expressed by the equations

\begin{equation}
\sigma_{ij}n_j=0\hspace{1cm}\mbox{at}\hspace{0.4cm} r=R
\hspace{0.4cm}\mbox{and at}\hspace{0.4cm}r=A\hspace{0.5cm}(R\geq a \geq 0),
\label{bc}
\end{equation}

where the sphere's surface $S\/$ has outward normal $\bf{n}$. The possibility
of a spherical shell ($a=R$), and that of a solid sphere ($a=0$), are
allowed. The stress tensor $\sigma_{ij}$ is given by \cite{ll}

\begin{equation}
\sigma_{ij}=\lambda\, u_{k,k}\, \delta_{ij}\,+\,2\,\mu \,u_{(i,j)}.            
\end{equation}

The general solution to (\ref{eigenv}) can be cast in the form

\begin{eqnarray}
{\bf u}({\bf x})&=&C_o\,\nabla\phi({\bf x})\,+\,i C_1\,{\bf L}\psi({\bf x})
\,+\,i C_2\,\nabla\times{\bf L}\psi({\bf x}) + \nonumber \\
&& D_o\,\nabla\tilde\phi({\bf x})\,+\,i D_1\,{\bf L}\tilde\psi({\bf x})
\,+\,i D_2\,\nabla\times{\bf L}\tilde\psi({\bf x}),
\end{eqnarray}

where $C_i$, $D_i$ are constants, ${\bf L}\equiv \,-i{\bf x}\times\nabla$, and
the scalar functions $\phi$, $\psi$, $\tilde\phi$, $\tilde\psi$ are given by

\begin{eqnarray}
\phi({\bf x})\,=\,j_l(qr)\,Y_{lm}(\theta,\varphi),&\hspace{1em}&
\psi({\bf x})\,=\,j_l(kr)\,Y_{lm}(\theta,\varphi), \\
\tilde\phi({\bf x})\,=\,y_l(qr)\,Y_{lm}(\theta,\varphi),&\hspace{1em}&
\tilde\psi({\bf x})\,=\,y_l(kr)\,Y_{lm}(\theta,\varphi),
\end{eqnarray}

where $q\equiv k\sqrt{\mu/(\lambda+\mu)}$, and $Y_{lm}\/$ denotes a
spherical harmonic. Finally, $j_l$ and $y_l$ are the standard Bessel functions
of the first and second kind, respectively (see, e.g., \cite{as72}). The
latter (which are singular at the origin) must be included in our case, as
$r=0$ lies outside the boundary $S\/$. The boundary conditions (\ref{bc})
become, after rather lengthy calculations, a system of linear equations
which splits up into a $4\times 4$ linear system for $(C_o,C_2,D_o,D_2)$,
and a $2\times 2$ system for $(C_1, D_1)$. That is, we have a linear 
system of the form:

\begin{equation}
\left(
\begin{array}{cc}
{\bf A_P}&0 \\ 0&{\bf A_T}
\end{array} \right) \left(
\begin{array}{c}
{\bf C_P}\\{\bf C_T}
\end{array} \right) = 0, \label{lsys}
\end{equation}

with

\begin{equation} 
{\bf C_P}\equiv (C_o,C_2,D_o,D_2)^t
,\hspace{3em} {\bf C_T}\equiv (C_1,D_1)^t,
\end{equation}
where the superscript $t$ denotes transpostion, and
the corresponding matrices are:
\begin{equation}
{\bf A_P}=\left(
\begin{array}{cccc}
\beta_4(qR)&-l(l+1)s^{-2}\beta_1(kR)&\tilde\beta_4(qR)&
-l(l+1)s^{-2}\tilde\beta_1(kR) \\
\beta_1(qR)&-s^{-2}\beta_3(kR)&\tilde\beta_1(qR)&
-s^2\tilde\beta_3(kR) \\
\beta_4(qa)&-l(l+1)s^{-2}\beta_1(ka)&\tilde\beta_4(qa)&
-l(l+1)s^{-2}\tilde\beta_1(ka) \\
\beta_1(qa)&-s^{-2}\beta_3(ka)&\tilde\beta_1(qa)&
-s^{-2}\tilde\beta_3(ka) \\
\end{array} \right) \label{AP}
\end{equation}

and

\begin{equation}
{\bf A_T}=\left(
\begin{array}{cc}
\beta_1(kR)&\tilde\beta_1(kR)\\
\beta_1(ka)&\tilde\beta_1(ka)
\end{array} \right) \label{AC}
\end{equation}

Here $s\equiv q/k$, and we have introduced the set of functions:

\begin{eqnarray}
&\beta_o(z)\equiv j_l(z)z^{-2}
\hspace{2em}\beta_1(z)\equiv\left( j_l(z)z^{-1}\right)'
\hspace{2em}\beta_2(z)\equiv j_l''(z)&\\
&\beta_3(z)\equiv\frac{1}{2}\beta_2(z)-\left\{1-\frac{l(l+1)}{2}\right\}
\beta_o(z),\hspace{1em}
\beta_4(z)\equiv\beta_2(z)-\frac{\lambda}{2\mu}z^2\beta_o(z) , &
\end{eqnarray}
while the tilded ones are their singular counterparts, with $y_l\/$ instead
of $j_l\/$ (i.e., $\tilde\beta_o(z)\equiv y_l(z)\,z^{-2}$, and so on). The 
matrices ${\bf A_P}$ and ${\bf A_T}$ are functions of $kR\/$, and depend on
the parameter $a/R\/$, and, in the case of ${\bf A_P}$, also on $s\/$
\footnote{This parameter is a function of the Poisson ratio $\sigma\/$;
for the usual value $\sigma=1/3$, $s=0.5$ and $\lambda/\mu=2$. These values
are assumed, unless otherwise stated.}.
The {\it discrete\/} set of $kR$ values that make compatible the system
(\ref{lsys}) constitute the {\it spectrum\/} of the elastic sphere. We can
distinguish two families of normal modes:

(i) {\em Toroidal modes}. These are characterized by

\begin{equation}
\det {\bf A_T}=0, \hspace{2em} {\bf C_P}=0. \label{modet}
\end{equation}

\begin{figure}
\psfig{file=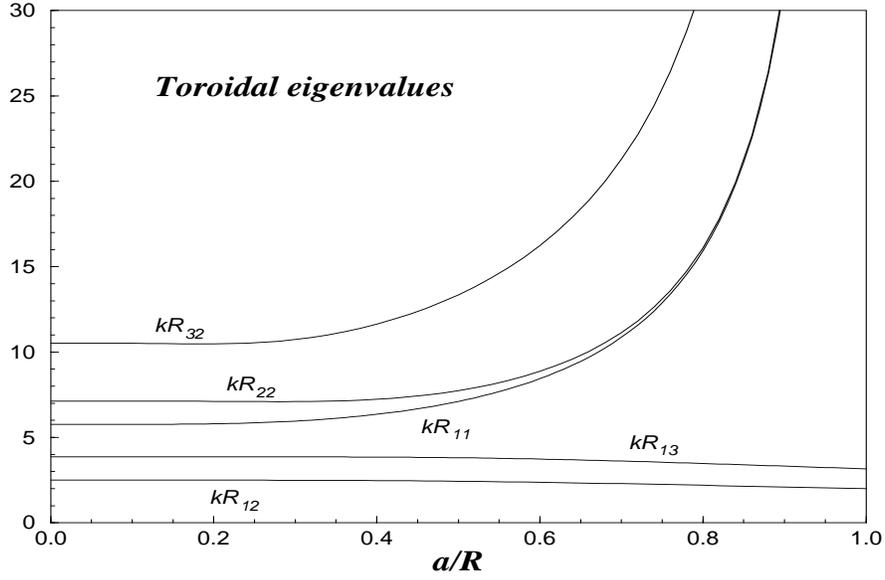,height=12cm,width=10cm,rheight=7.8cm,bbllx=-4cm,bblly=-1.4cm,bburx=13.2cm,bbury=23.6cm}
\caption{Functional dependence of the first few toroidal eigenvalues of a
hollow sphere on the ratio $a/R\/$. Solid sphere values ($a\/$\,=\,0) are
found on the intersections with the ordinate axis.  \label{1}}
\end{figure}

\begin{figure}
\psfig{file=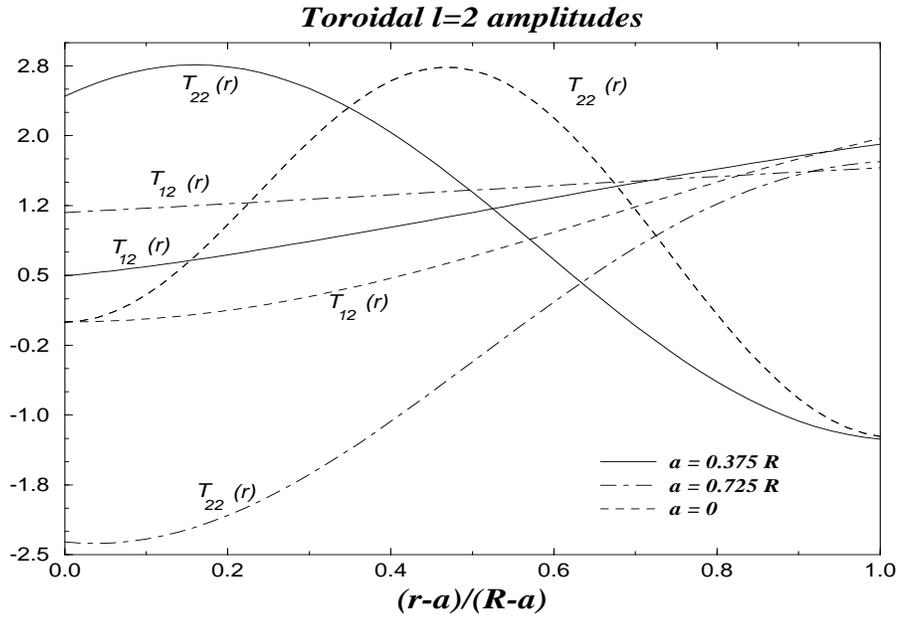,height=12cm,width=10cm,rheight=7.8cm,bbllx=-4cm,bblly=-1.4cm,bburx=13.2cm,bbury=23.6cm}
\caption{Toroidal mode radial functions for the first two quadrupole harmonics
and a few values of the geometric ratio $a/R\/$. The magnitude represented in
abscissas is such that the region plotted spans radially the material
thickness of the hollow sphere.  \label{2}}
\end{figure}

Hence they are purely tangential, and their frequencies depend only on the
ratio $a/R$. Their amplitudes are

\begin{equation}
{\bf u}_{nlm}^{T}({\bf x})=T_{nl}(r)\,i{\bf L}Y_{lm}(\theta,\phi),
\end{equation}

with

\begin{equation}
T_{nl}(r)=C_1(n,l)\left\{\tilde\beta_1(k_{nl}R)j_l(k_{nl}r)-
                         \beta_1(k_{nl}R)y_l(k_{nl}r)\right\},
\end{equation}
where $C_1(n,l)$ is fixed by the chosen normalization. The corresponding
eigenvalues are obtained as solutions to the trascendental equation
(\ref{modet}). For the degenerate limit $a=R$ the equation to be solved is

\begin{equation}
\det \left(
\begin{array}{cc}
\beta_1(kR)&\tilde\beta_1(kR)\\
\beta_1'(kR)&\tilde\beta_1'(kR)
\end{array}
\right) = 0,
\end{equation}
with the prime denoting  differentation respect to the argument. Using
standard properties of Bessel functions \cite{as72}, it can be easily
shown that

\[\beta_1(kR) \tilde\beta_1'(kR)-\tilde\beta_1(kR)\beta_1'(kR)=(kR)^{-6}[(kR)^2+
2-l(l+1)], \]
and, in this case, there is only {\em one} eigenvalue for each $l>1$, given
by the only root of the above equation, $(k_lR)^2=l(l+1)-2$\ \footnote{
This equation shows explicitly a property shared by all toroidal modes,
namely, that their dimensionless eigenvalues $k_{nl}R\/$ do {\it not\/}
depend on the elastic properties of the material.}.
Figure \ref{1} displays $k_{nl}R\/$ as a function of $a/R\/$ for the first
few toroidal modes. The existence of just one mode for each $l>1$ in the thin
shell limit shows as a divergence of $k_{nl}R\/$ when $a/R$ approaches 1 and
$n>1$. In figure \ref{2} we plot the normalized toroidal amplitudes
$T_{nl}(r)$ for two quadrupolar modes and three different values of the
parameter $a/R$. We observe that their absolute values at the outer surface
show little dependence on the ratio $a/R$.


(ii) {\em Spheroidal modes}.
This second family is charaterized by:

\begin{equation}
\det{\bf A_P}=0, \hspace{2em} {\bf C_T}=0. \label{modesp}
\end{equation}

In this case, the expressions get more involved, as we have to handle
a $4\times 4$ determinant. Once the spectrum $k_{nl}$ is found for given
$a/R$ and $s$, the system (\ref{lsys}) can be solved for $C_2/C_o$, $D_o/C_o$,
and $D_2/C_o$. If we label these coefficients $p_o(n,l)$, $p_1(n,l)$,
$p_2(n,l)$, the eigenmodes can be written as

\begin{equation}
{\bf u}_{nlm}^P= N_{nl}(r) Y_{lm}(\theta,\phi){\bf n}-iE_{nl}(r){\bf n}\times
{\bf L} Y_{lm}(\theta,\phi), 
\end{equation}

with

\begin{eqnarray}
 N_{nl}(r)&=& 
  C_o(n,l)\left[j_l'(q_{nl}r)-p_o(n,l) \frac{l(l+1)}{q_{nl}r}\,
  j_l(k_{nl}r)+\right.\nonumber \\
 && \left. p_1(n,l)\, y_l'(q_{nl}r)-p_2(n,l) \frac{l(l+1)}{q_{nl}r}\,
  y_l(k_{nl}r)\right], \label{N}\\
 E_ {nl}(r)&=& C_o(n,l)\frac{1}{q_{nl}r}\left[ j_l(q_{nl}r)-
  p_o(n,l)\{k_{nl}r\,j_l(k_{nl}r)\}'+ \right. \nonumber\\
  && \left. p_1(n,l)y_l(q_{nl}r)-
  p_2(n,l)\{k_{nl}r\,y_l(k_{nl}r)\}'\right], \label{E}
\end{eqnarray}
where $C_o(n,l)$ is, again, free up to normalization. The spectrum for the
degenerate case $a=R$ is given by the solutions to

\begin{equation}
\det \left(
\begin{array}{cccc}
\beta_4(qR)&-l(l+1)s^{-2}\beta_1(kR)&\tilde\beta_4(qR)&
-l(l+1)s^{-2}\tilde\beta_1(kR) \\
\beta_1(qR)&-s^{-2}\beta_3(kR)&\tilde\beta_1(qR)&
-s^2\tilde\beta_3(kR) \\
\beta_4'(qR)&-l(l+1)s^{-1}\beta_1'(kR)&\tilde\beta_4'(qR)&
-l(l+1)s^{-1}\tilde\beta_1'(kR) \\
\beta_1'(qR)&-s^{-1}\beta_3'(kR)&\tilde\beta_1'(qR)&
-s^{-1}\tilde\beta_3'(kR) 
\end{array} \right) = 0, \label{shell}
\end{equation}
which happens to have {\em two\/} solutions for each value of $l\/$ when
$l>1$ and only {\em one\/} root for $l<2$\ \footnote{The purely radial case
$l=0$ is simpler, because the eigenvalue equation (protect\ref{shell}) becomes
\[\beta_4(qR)\tilde\beta_4'(qR)-\beta_4'(qR)\tilde\beta_4(qR)=0,\]
and has only one solution, namely $qR=(\mu/\lambda)\sqrt{3-\mu/\lambda}$.
Unlike toroidal eigenvalues, spheroidal ones do depend on $\mu/\lambda$.}.

\begin{figure}
\psfig{file=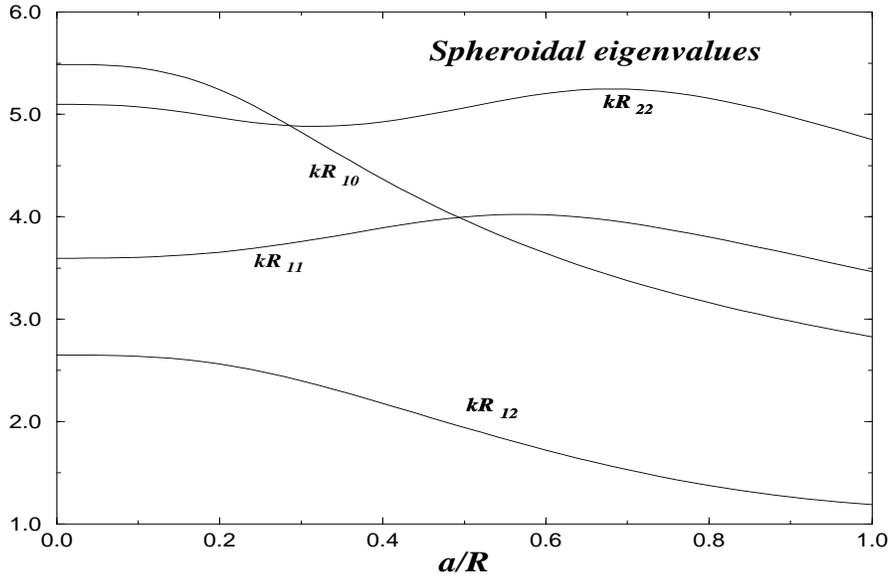,height=12cm,width=10cm,rheight=7.8cm,bbllx=-4cm,bblly=-1.4cm,bburx=13.2cm,bbury=23.6cm}
\caption{Functional dependence of the first few spheroidal eigenvalues of a
hollow sphere on the ratio $a/R\/$. Solid sphere values ($a\/$\,=\,0) are
found on the intersections with the ordinate axis.  \label{3}}
\end{figure}

\begin{figure}
\psfig{file=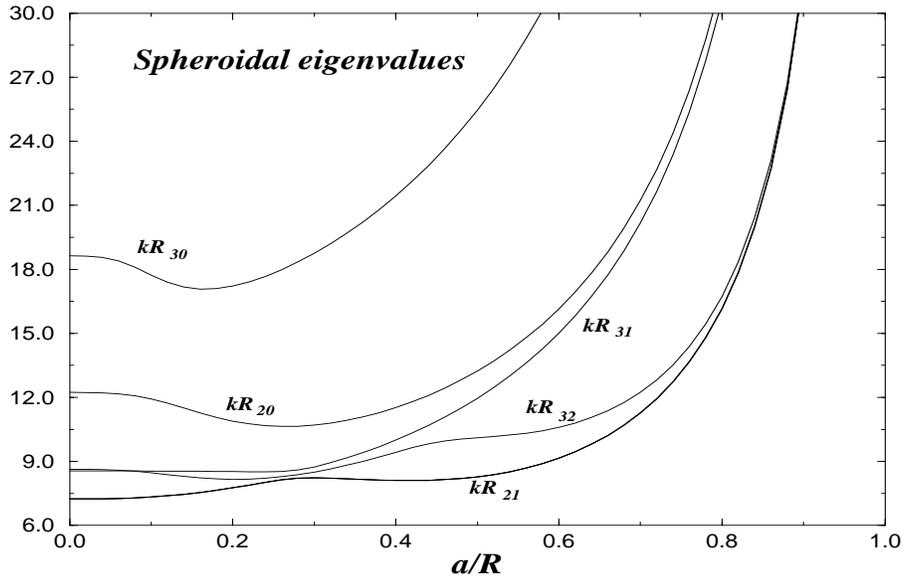,height=12cm,width=10cm,rheight=7.8cm,bbllx=-4cm,bblly=-1.4cm,bburx=13.2cm,bbury=23.6cm}
\caption{Functional dependence of higher spheroidal eigenvalues of a hollow
sphere on the ratio $a/R\/$. The harmonics in this graph do not exist in the
thin shell limit, and this shows as divergencies as $a\/$ approaches $R\/$.
\label{4}}
\end{figure}

\begin{figure}
\psfig{file=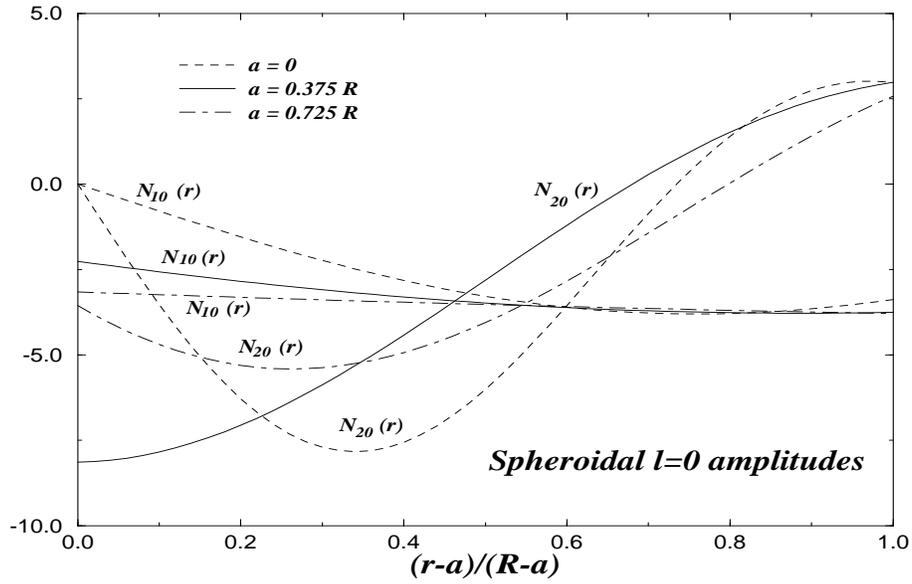,height=12cm,width=10cm,rheight=7.8cm,bbllx=-4cm,bblly=-1.4cm,bburx=13.2cm,bbury=23.6cm}
\caption{Spheroidal mode radial $N\/$-functions ---see equation
(\protect\ref{N})--- for the first two monopole harmonics and a few values of
the geometric ratio $a/R\/$. The magnitude represented in abscissas is such
that the region plotted spans radially the material thickness of the hollow
sphere.   \label{5}}
\end{figure}

\begin{figure}
\psfig{file=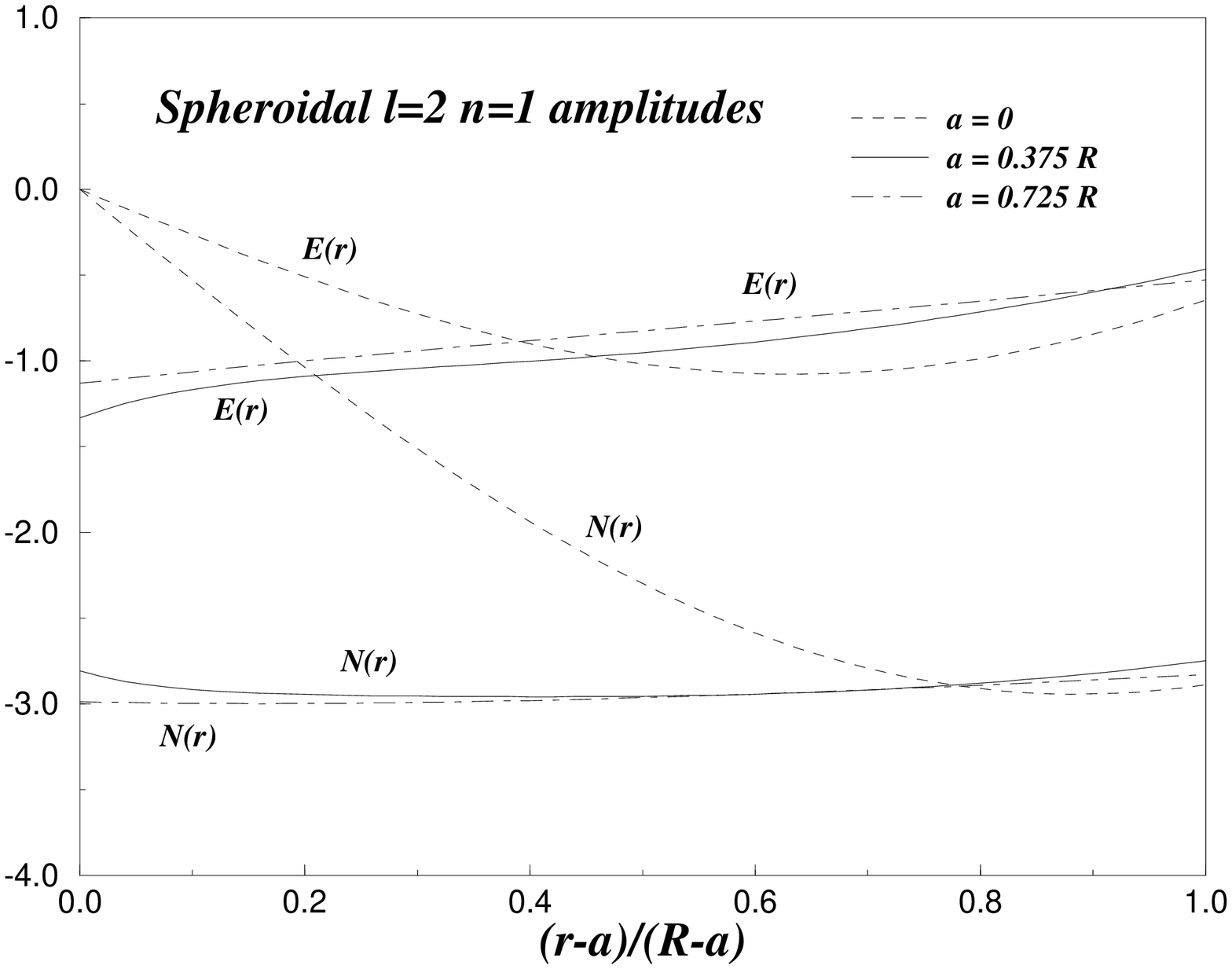,height=12cm,width=10cm,rheight=7.8cm,bbllx=-4cm,bblly=-1.4cm,bburx=13.2cm,bbury=23.6cm}
\caption{Spheroidal mode radial functions ---see equations (\protect\ref{N})
and (\protect\ref{E})--- for the {\it first\/} quadrupole harmonic.
\label{6}}
\end{figure}

\begin{figure}
\psfig{file=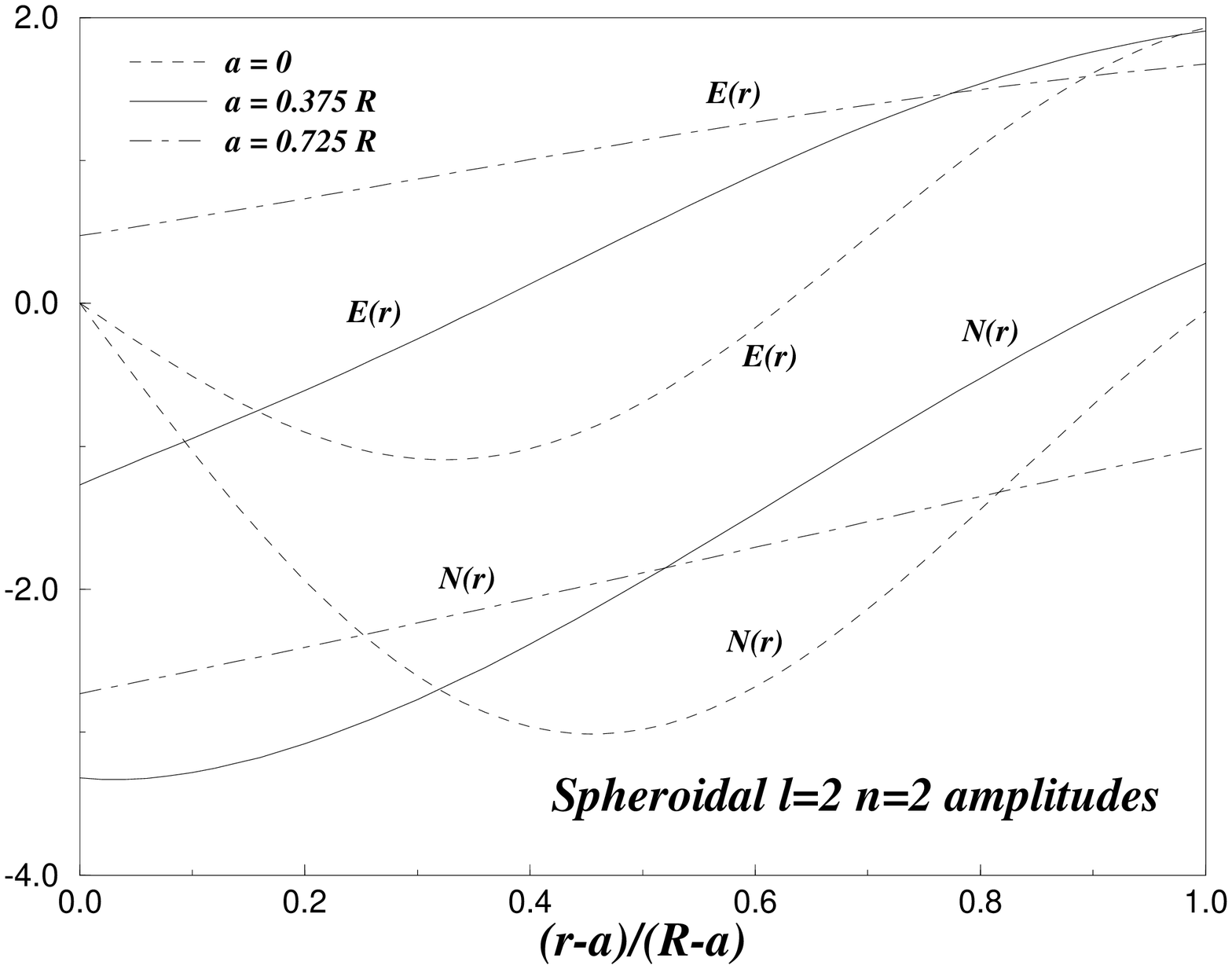,height=12cm,width=10cm,rheight=7.8cm,bbllx=-4cm,bblly=-1.4cm,bburx=13.2cm,bbury=23.6cm}
\caption{Spheroidal mode radial functions ---see equations (\protect\ref{N})
and (\protect\ref{E})--- for the {\it second\/} quadrupole harmonic.
\label{7}}
\end{figure}

Plotting $k_{nl}R\/$ as a function of $a/R$, we see that the third and higher
roots diverge as the inner radius approaches $R\/$, see figures \ref{3} and
\ref{4}. Figures \ref{5}--\ref{7} show the normalized radial functions for
a few spheroidal modes and values of $a/R$. As in the toroidal case, their
values at $r=R\/$ (where measurements using transducers are to be made
eventually) are nearly independent of $a/R$.


\section{Cross section for the hollow sphere}

A convenient way to characterise a resonant detector sensitivity is through
its GW energy absorption cross section, defined as

\begin{equation}
  \sigma_{\rm abs}(\omega)=\frac{\Delta E_a(\omega)}{\Phi(\omega)}
  \label{100}
\end{equation}
where $\Delta E_a(\omega)$ is the energy absorbed by the detector at
frequency $\omega$, and $\Phi(\omega)$ is the incident flux density
expressed e.g. in watt/m$^2$\,Hz. Estimation of $\sigma_{\rm abs}(\omega)$
requires a hypothesis about the underlying gravitation theory to calculate
$\Phi(\omega)$, and specification of the antenna's geometry to calculate
$\Delta E_a(\omega)$. Here we shall assume that General Relativity is the
correct gravitation theory, and proceed to calculate the oscillation energy
of the solid as a consequence of its excitation by an incoming GW, which we
shall naturally identify with $\Delta E_a(\omega)$. We briefly sketch the
details of the process now.

As shown in \cite{ll}, an elastic solid's response to a GW force can be
expressed by a very general formula, which is easily particularised to a
spherically symmetric body such as the {\it solid\/} sphere {\it or\/} the
{\it hollow\/} sphere. In both cases, as we have just seen, the vibration
eigenmodes belong into two families (spheroidal and toroidal), but GWs only
couple to {\it quadrupole spheroidal harmonics\/}. If the frequencies of
these modes are noted by $\omega_{n2}$ ($n\/$\,=\,1 for the lowest value,
$n\/$\,=\,2 for the next, etc.) and the corresponding wavefunctions by
${\bf u}_{n2m}({\bf x})$ then the elastic displacements are given by

\begin{equation}
   {\bf u}({\bf x},t) = \sum_{n=1}^\infty\,\frac{b_n}{\omega_{n2}}\,\left[
   \sum_{m=-2}^2\,{\bf u}_{n2m}({\bf x})\,g_{n2}^{(m)}(t)\right]
   \label{3.16}
\end{equation}
where 

\begin{equation}
 g_{nl}^{(m)}(t) \equiv \int_0^t g^{(m)}(t')\,\sin\omega_{n2}(t-t')\,dt'
 \ ,\qquad (m=-2,\ldots,2)     \label{3.16b}
\end{equation}

and $g^{(m)}(t)$ are the quadrupole components of the Riemann tensor,
while $b_n\/$ is an {\it overlapping integral factor\/} of the GW's tidal
coefficient over the solid's extension. Much like in the case of a solid
sphere, it has dimensions of length, and is given by a definite integral
of the radial terms in the wavefunction ${\bf u}_{n2m}({\bf x})$; more
specifically,

\begin{equation}
  \frac{b_n}{R} = -\frac{\rho}{M}\int_a^R r^3 [N_{n2}(r)+3E_{n2}(r)]\,dr =
  -\frac{C_o(n,2)}{4\pi\,q_{n2}R}[G_2(R)-G_2(a)],
\end{equation}
where we have introduced the dimensionless function

\begin{equation}
  G_2(z) \equiv \frac{z^3}{R^3-a^3} \left[j_2(q_{n2}z)+p_1(n,2)y_2(q_{n2}z) -
  - 3 p_o(n,2)j_2(k_{n2}z)-3 p_2(n,2)y_2(k_{n2}z) \right]
\end{equation}
and assumed the following normalization for the wavefunctions:

\begin{equation}
  \int_{\rm Solid}\left|{\bf u}_{nlm}\right|^2\,\rho\,d^3{\bf x} =
  \int_a^R r^2\,dr\,\rho\,\left[ N_{ln}^2(r)+l(l+1) E_{ln}^2(r)\right] = M
\end{equation}

The calculation of $\Delta E_a(\omega)$ can now be pursued along the lines
set up in reference \cite{ll}: the Fourier transform ${\bf U}({\bf x},\omega)$
of the response function ${\bf u}({\bf x},t)$ of equation (\ref{3.16}) is
calculated, whereby the {\it spectral energy density\/} can be obtained as

\begin{equation}
  W(\omega) = \frac{1}{T}\;\int_{\rm Solid}\frac{1}{2}\,\omega^2\,
        \left|{\bf U}({\bf x},\omega)\right|^2\,\rho\,d^3x
\end{equation}
where $T\/$ is the {\it integration time\/} of the signal in the detector.
The energy deposited by the GW in the $n\/$-th quadrupole mode is hence
calculated by integration of this spectral density over the linewidth of
the mode. It is readily found that

\begin{equation}
  \Delta E_a(\omega_{n2}) = \frac{1}{2}\,M\,b_n^2\,
  \sum_{m=-2}^2\,\left|G^{(m)}(\omega_{n2})\right|^2    \label{4.10b}
\end{equation}
where $G^{(m)}(\omega)$ is the Fourier transform of $g^{(m)}(t)$.

The GW flux in the denominator of (\ref{100}) is (clearly) proportional to
the sum in the rhs of (\ref{4.10b}), the proportionality factor being in turn
proportional to $\omega^2$ ---see \cite{ll} for a detailed discussion---,
so we finally obtain

\begin{equation}
  \sigma_n \equiv
  \sigma_{\rm abs}(\omega_{n2})= \frac{16\pi^2}{15}\,\frac{GMv_t^2}{c^3}
  (k_{n2}b_{n})^2 \label{sigma}
\end{equation}

where $v_t^2\/$\,=\,$\mu/\rho$, $M\/$ is the detector's mass, and $G\/$ is
Newton's constant. This equation allows relatively easy numerical evaluation
of the cross sections, as well defined computer programmes can be written for
the purpose.

\begin{figure}
\psfig{file=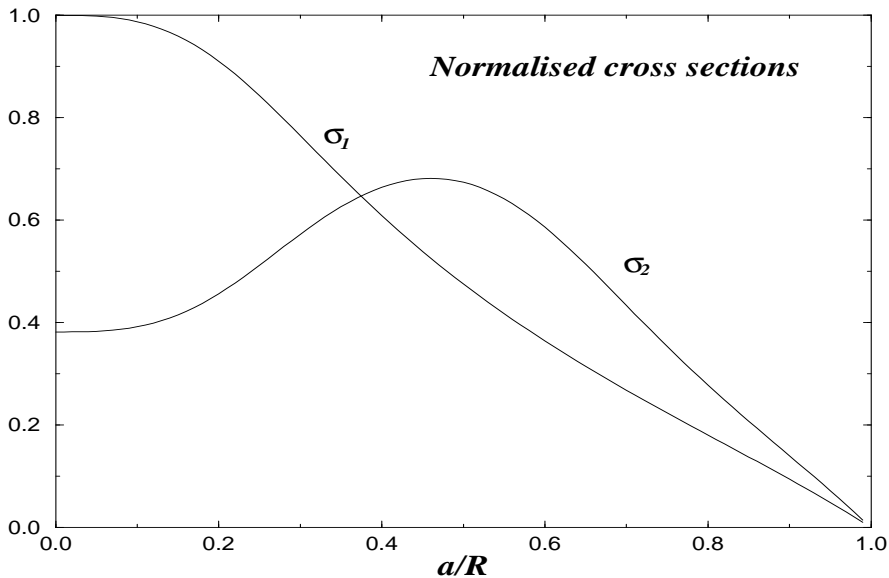,height=12cm,width=10cm,rheight=7.8cm,bbllx=-4cm,bblly=-1.4cm,bburx=13.2cm,bbury=23.6cm}
\caption{Cross sections of a hollow sphere in its first two quadrupole modes
as a function of thickness. Values are referred to the cross section of a
solid sphere in its first quadrupole resonance, whose {\it radius\/} is
assumed to be equal to the {\it outer\/} radius of the hollow sphere.
\label{8}}
\end{figure}

\begin{figure}
\psfig{file=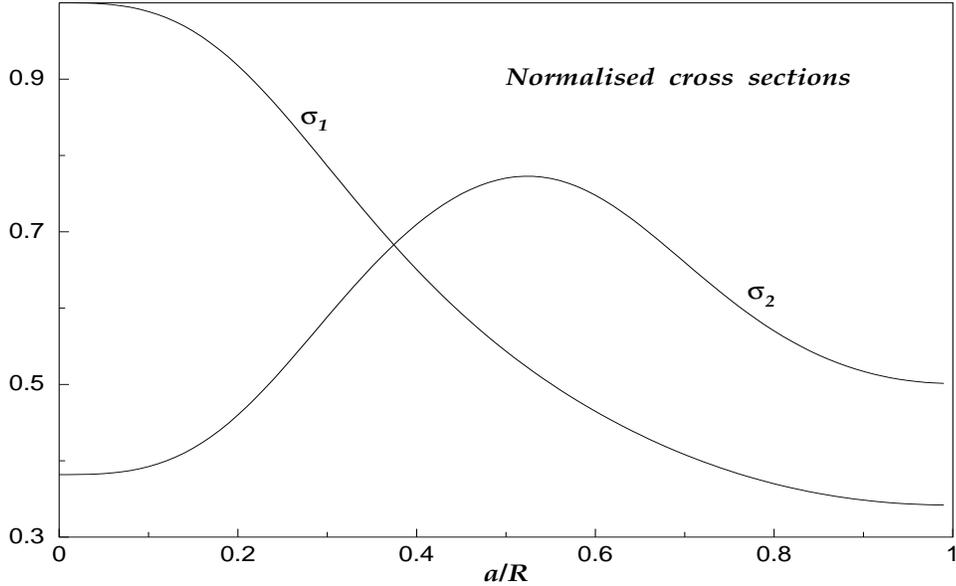,height=12cm,width=10cm,rheight=7.8cm,bbllx=-3.5cm,bblly=-1.4cm,bburx=14.2cm,bbury=23.6cm}
\caption{Cross sections of a hollow sphere in its first two quadrupole modes
as a function of thickness. Values are referred to the cross section of a
solid sphere in its first quadrupole resonance, whose {\it mass\/} equals
that of the hollow sphere.
\label{8a}}
\end{figure}

As we have seen in section 2 above, the eigenvalues and wavefunctions of a
hollow sphere only depend on the ratio $a/R\/$, and therefore so does the
quantity $(k_{n2}b_n)$ in (\ref{sigma}). So the cross section $\sigma_n\/$
only depends on that ratio, too, once a suitable unit of {\it mass\/} is
adopted for reference. In figures \ref{8} and \ref{8a} we plot $\sigma_n$
for the first two quadrupole modes of the hollow sphere in two different
circumstances: in figure \ref{8} we assume a hollow sphere of {\it fixed\/}
outer radius ---thus its mass decreases with thickness---, and in figure
\ref{8a} we have instead assumed that the mass of the hollow sphere is
{\it fixed\/}, so that its geometrical size increases as it gets thinner.
In either case we see that, for the higher mode, the maximum cross section
does not happen at $a\/$=0, but at some intermediate inner radius: for
$a\approx 0.37745 R$, the cross-section for the second quadrupole mode
equals that of the first, and we have the possibility of working with
a detector {\em with the same (high) sensitivity at two frequencies.}

\section{Sensitivity to GW signals}

We assume that the mechanical oscillations induced in a resonant mass by
the interaction with the GW are transformed into electrical signals by a set
of identical noiseless transducers (for the sake of simplicity, we consider
here non-resonant transducers), perfectly matched to electronic amplifiers
with noise temperature $T_n\/$. Unavoidably, Brownian motion noise associated
with dissipation in the antenna and electronic noise from the amplifiers
limit the sensitivity of the detector. We refer the reader
to~\cite{giffard,pz,thorne} for a complete discussion on the sensitivity
of resonant-mass detectors and report here only a few basic formulas for
the evaluation of the detector sensitivity to various signals.

The total noise at the output of each resonant mode
can be seen as due to an input noise generator having spectral
density of strain $S_h(f)$, acting on a noiseless oscillator. 
$S_{h}(f)$ represents the input GW spectrum that would produce a signal
equal to the noise spectrum actually observed at the output of the
detector instrumentation. In a resonant-mass detector, 
this function is a resonant curve and can
be characterized by its value at resonance $S_{h}(f_n)$ 
and by its half height width. $S_{h}(f_n)$ can be written as:

\begin{equation}
  S_h(f_n) = \frac{G}{c^3}\frac{4kT_e}{\sigma_{n} Q_{n} f_{n}}
\label{esseacca}
\end{equation}            

Here $T_e\/$ is the thermodynamic temperature of the detector plus a
back-action contribution from the amplifiers, and $Q_n\/$ is the quality
factor of the mode.

The half height width of $S_{h}(f)$ gives the bandwidth of the resonant
mode:

\begin{equation}
\Delta f_n = \frac{f_n}{Q_n} \Gamma_{n} ^{-1/2}
\label{deltaeffe}
\end{equation}

Here, $\Gamma_n$ is the ratio of the wideband noise in the $n\/$-th resonance
bandwidth to the narrowband noise,

\begin{equation}
\Gamma_n \simeq \frac{T_n}{2\beta_n\,Q_n\,T_e},
\end{equation}

where $\beta_n$ is the transducer coupling factor, defined as the fraction of
the total mode energy available at the transducer output.

In practice $\Gamma_n$\,$\ll$\,1 and the bandwidth is much larger than the
pure resonance linewidth $f_{n}/Q_{n}$. In the limit
$\Gamma_{n}\rightarrow 0$, the bandwidth becomes infinite. The bandwidth of
the present resonant bars is of the order of a few Hz~\cite{mr}. If a
quantum limited readout system were available, values of the order of
100 Hz could be reached~\cite{paik,bill}.

The equations (\ref{esseacca} and \ref{deltaeffe}) can be used to 
characterize the sensitivity of the
quadrupole modes of a hollow spherical resonant-mass detector. The optimum
performance is obtained by filtering the output with a filter matched to
the signal. The energy signal-to-noise ratio ($SNR$) of the filter output
is given by the well-known formula

\begin{equation}
SNR = \int^{+\infty}_{-\infty}\frac{|H(f)|^2} {S_{h}(f)}\,df
\label{ev5}
\end{equation}

where $H(f)$ is the Fourier transform of $h(t)$.

We now report the {\it SNR\/} of a hollow spherical detector for various GW
signals. To be specific, we shall assume that the thermodynamic temperature
of the detector can be reduced to below 50 mK, and that the quality factors
of the modes are of the order of 10$^7$, so that the overall detector noise
will be dominated by the electronic amplifier noise. If we express the energy
of the latter as a multiple of the {\it quantum limit\/},
i.e., $kT_n\/$\,=\,$N\hbar\omega$ then the strain spectral density becomes

\begin{equation}
  S_h(f_n) \simeq \frac{G}{c^3}\,\frac{4\pi\beta_nN\hbar}{\sigma_n}
\end{equation}

In these conditions the fractional bandwidth $\Delta f_n/f_n\/$ becomes of the
order of $\beta_n$ that we assume of about 0.1. We shall consider hollow 
spheres made of the usual aluminium alloy Al\,5056 and of a recently 
investigated copper alloy (CuAl) \cite{giorgetto}. Table \ref{t1} displays 
numerical values of the most relevant
parameters for a few example detectors with a noise level equal to the 
quantum limit, i.e. $N$\,=\,1.

\vskip 1.0cm

\begin{table}
\caption{Main features and sensitivities for several hypothetical hollow
spheres of two different materials.   \label{t1}}
\begin{tabular}{cccccccccc}
 & M (ton) & $2R$ (m) & $(R-a)$ (cm) & $f_1$ (Hz) & $f_2$ (Hz)
 & $\sigma_1$ (m$^2$ Hz) & $\sigma_2$ (m$^2$ Hz) & $\sqrt{S_{h1}}
 \left(Hz^{-1/2}\right)$ & $\sqrt{S_{h2}} \left(Hz^{-1/2}\right)$ \\ \hline 
CuAl   & 200 & 4 & 81 & 395 & 1188 & 1.5 $10^{-23}$ 
       & 2.4 $10^{-23}$ & 4.7 $10^{-24}$ &3.7 $10^{-24}$ \\
       & 200 & 6 & 25 & 191 & 753  & 1.1 $10^{-23}$
       & 1.7 $10^{-23}$ & 5.4 $10^{-24}$ & 4.4 $10^{-24}$ \\
       & 100 & 4 & 31 & 302 & 1161 & 5.8 $10^{-24}$
       & 8.8 $10^{-24}$ & 7.5 $10^{-24}$ & 6.1 $10^{-24}$ \\
       & 100 & 6 & 12 & 185 & 738  & 5.6 $10^{-24}$
       & 8.2 $10^{-24}$ & 7.7 $10^{-24}$ & 6.3 $10^{-24}$ \\ 
       & 40  & 3 & 22 & 399 & 1543 & 2.3 $10^{-24}$
       & 3.5 $10^{-24}$ & 1.2 $10^{-23}$ & 9.7 $10^{-24}$ \\ 
       & 40  & 4 & 11 & 281 & 1115 & 2.2 $10^{-24}$
       & 3.3 $10^{-24}$ & 1.2 $10^{-23}$ & 1.0 $10^{-23}$ \\ 
Al5056 & 200 & 6 & 90 & 273 & 935  & 1.8 $10^{-23}$
       & 2.9 $10^{-23}$ & 4.3 $10^{-24}$ & 3.4 $10^{-24}$ \\ 
       & 100 & 6 & 37 & 230 & 896  & 7.7 $10^{-24}$
       & 1.2 $10^{-23}$ & 6.6 $10^{-24}$ & 5.4 $10^{-24}$ \\ 
       & 40  & 4 & 35 & 361 & 1370 & 3.2 $10^{-24}$
       & 4.8 $10^{-24}$ & 1.0 $10^{-23}$ & 8.3 $10^{-24}$ \\ 
       & 40  & 6 & 14 & 218 & 866  & 3.0 $10^{-24}$
       & 4.4 $10^{-24}$ & 1.0 $10^{-23}$ & 8.7 $10^{-24}$ \\ 
\end{tabular}
\end{table}

\subsection{Bursts}

We model the burst signal as a featureless waveform, rising quickly to an 
amplitude $h_{0}$ and lasting for a time $\tau_{g}$ much shorter than the
detector integration time $\Delta t = \Delta f_{n}^{-1}$. Its Fourier
transform will be considered constant within the detector bandwidth: $H(f)
\simeq H(f_{n}) = H_{0}$. From (\ref{ev5}) we get

\begin{equation}
SNR = \frac{2\pi\Delta f_{n} H_{0}^{2}}{S_{h}(f_n)}
\label{ev6}
\end{equation}

For {\it SNR\/}\,=\,1, and using the equation
$H_{0}^{\rm min}$ = $h_{0}^{\rm min}\tau_{g}$, we find

\begin{equation}
\left(h_0^{\rm min}\right)_{\rm burst} =
\tau_{g}^{-1}\left[\frac{S_{h}(f_{n})}{2\pi \Delta f_n}\right]^{1/2}
\end{equation}

The level $h_0^{min}\simeq 10^{-22}$ can be reached by the lowest order mode
of a typical large hollow spherical detector such as the one being considered.
The GW luminosity of burst sources is still largely unknown, so it is
difficult to accurately estimate their detectability. The above sensitivity
is however likely to enable the detection of GW collapses in the Virgo
cluster for an energy conversion of 10$^{-4}$\,$M_{\odot}$ into a millisecond
GW burst. See table \ref{t2} for a few specific examples.

\subsection{Monochromatic signals}

We consider a sinusoidal wave of amplitude $h_{0}$ and frequency $f_s$
constant over the observation time $t_m$. The Fourier transform amplitude at
$f_n$ is $\frac{1}{2}h_{0}t_m$ with a bandwidth given by $t_m^{-1}$.
The {\it SNR\/} can be written as

\begin{equation}
  SNR = \frac{t_m\,h_0^2}{2 S_h(f_n)}\,\left\{1 +
  \Gamma_n\,\left[Q^2\left(1 - f_s^2/f_n^2\right) + f_s^2/f_n^2\right]\right\}
\end{equation}

For{\it SNR\/}\,=\,1 we obtain a minimum detectable value of $h_{0}$, which at
$f_{s}$\,=\,$f_{n}$ is

\begin{equation}
\left(h_{0}^{\rm min}\right)_{\rm m} =
\left[\frac{2S_{h}(f_{n})}{t_{m}}\right]^{1/2}
\end{equation}

See table \ref{t2} for a few specific examples. For instance, the nearby
pulsar~\cite{ph} PSR\,J0437-4715, at a distance of 150 pc, might emit at
347 Hz a GW amplitude (optimistically) of $2\times 10^{-26}$. This would
give {\it SNR\/}\,=\,100 on a hollow spherical detector having $M=100$ tons
after integrating the signal for 1 year.

\begin{table}
\caption{Sensitivity to burst and monochromatic (integrated for one year) GW
signals of a few hollow spheres of two different materials.   \label{t2}}
\begin{tabular}{cccccccccc}
 & M (ton) & $2R$ (m) & $(R-a)$ (cm) & $f_1$ (Hz) & $f_2$ (Hz)
 & $\left(h_{01}^{\rm min}\right)_{\rm burst}$ 
 & $\left(h_{02}^{\rm min}\right)_{\rm burst}$ 
 & $\left(h_{01}^{\rm min}\right)_{\rm m}$
 & $\left(h_{02}^{\rm min}\right)_{\rm m}$ \\[1 ex] \hline
CuAl   & 200 & 6 & 25 & 191 & 753 & 4.9 $10^{-22}$ & 2.0 $10^{-22}$
       & 1.4 $10^{-27}$ & 1.1 $10^{-27}$ \\
       & 100 & 4 & 31 & 302 & 1161 & 5.5 $10^{-22}$ & 2.3 $10^{-22}$
       & 1.9 $10^{-27}$ & 1.5 $10^{-27}$ \\
Al5056 & 100 & 6 & 37 & 230 & 896 & 5.5 $10^{-22}$ & 2.3 $10^{-22}$
       & 1.7 $10^{-27}$ & 1.4 $10^{-27}$ \\
\end{tabular}
\end{table}

\subsection{Chirps}

We consider here the interaction of the hollow spherical detector with the
waveform emitted by a binary system, consisting of either neutron stars or
black holes, in the inspiral phase. The system, in the Newtonian regime, has
a clean analytic behaviour, and emits a waveform of increasing amplitude
and frequency that can sweep up to the kHz range of frequency.

From the resonant-mass detector viewpoint, the chirp signal can be treated as
a transient GW, depositing energy in a time-scale short with respect to the
detector damping time \cite{bd}. We can then use (\ref{ev6}) to evaluate the
{\it SNR\/}, where the Fourier transform $H(f_n)$ at the resonant frequency
$f_n$ can be explicitly written as 

\begin{equation}
H(f_n) = \left\{\left[\int h(t)\cos(2\pi f_n t)\,dt\right]^2 +
		\left[\int h(t)\sin(2\pi f_n t)\,dt\right]^2\right\}^{1/2}
\label{ev9}
\end{equation}

with $h(t)$ indicating $h_+(t)$ or $h_{\times}(t)$. Substituting into
(\ref{ev9}) the well-known chirp waveforms for an optimally oriented orbit
of zero eccentricity in the Newtonian approximation~\cite{thorne}, the 
{\it SNR\/} for chirp detection is~\cite{sv}:

\begin{equation}
SNR = \frac{2^{1/3}5}{12} \frac{G^{5/3}}{c^3} \frac{\pi^2 \Delta f_n}{S_h(f_n)}
\frac{1}{r^2}
M_{c}^{5/3} (2\pi f_n)^{-7/3}
\label{ev7}
\end{equation}

$M_c$ is the chirp mass defined as
$M_c = (m_1 m_2)^{3/5}(m_1 + m_2)^{-1/5}$, where $m_1$ and $m_2$
are the masses of the two compact objects and $r\/$ is the distance 
to the source. The chirp mass is the only parameter that determines 
the frequency sweep rate of the chirp signal in the Newtonian approximation, 
and can be determined by a {\it double passage\/} technique \cite{sv}: 
much like in a solid sphere detector, one can measure the time delay 
$\tau_{2}-\tau_{1}$ between excitations of the first and second quadrupole 
modes on a hollow spherical detector to calculate the chirp mass through 
equation

\begin{equation}
M_c = 2^{8/5}\left(\frac{5}{256}\right)^{3/5}\,\frac{c^3}{G}\,
\left(\frac{\omega_2^{-8/3}-\omega_1^{-8/3}}{\tau_2-\tau_1}\right)^{3/5}
\label{viv1}
\end{equation}  

where $\omega_{1}$ and $\omega_{2}$ are the angular frequencies of the
first and second quadrupole modes, respectively. Time delays are of the 
order of a fraction of a second for the hollow spheres considered
in this paper, well within the timing possibilities of resonant mass detectors
\cite{vitatime}

\begin{figure}
\psfig{file=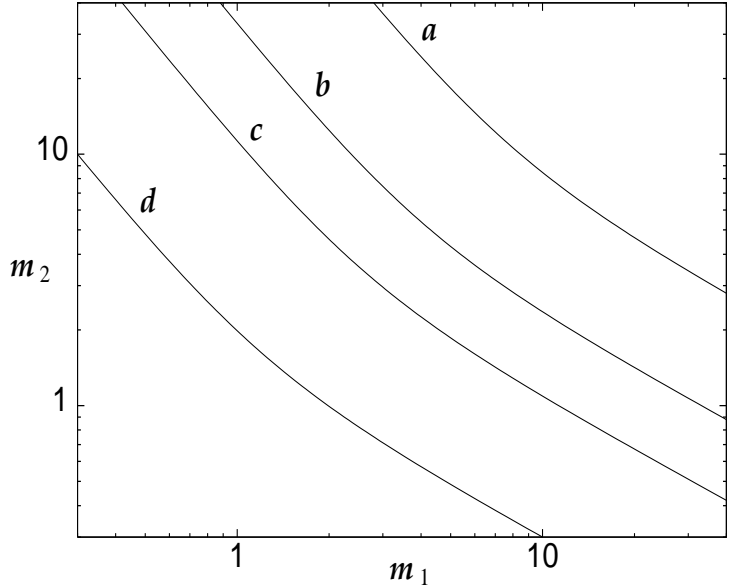,height=12cm,width=10cm,rheight=7.8cm,bbllx=-6cm,bblly=-1.4cm,bburx=11.2cm,bbury=23.6cm}
\caption{Contours of constant chirp mass $M_c$ in $m_1$, $m_2$ space. At each 
chirp mass corresponds the maximum distance $r$ at which the chirp can be 
observed with a SNR=10 by a 200 ton CuAl hollow sphere, 6 meters in diameter,
at its first resonance frequency $f_1 = 191$ Hz. The reported chirp mass
values (in units of solar masses) and the corresponding maximum distances 
are: $a: M_c = 8.0$, $r=214$ Mpc, $b: M_c = 4.0$, $r=119$ Mpc, $c: M_c =
2.6$, $r= 84$ Mpc, $d: M_c =1.2$, $r =45$ Mpc.}  
\label{9}
\end{figure}

\begin{figure}
\psfig{file=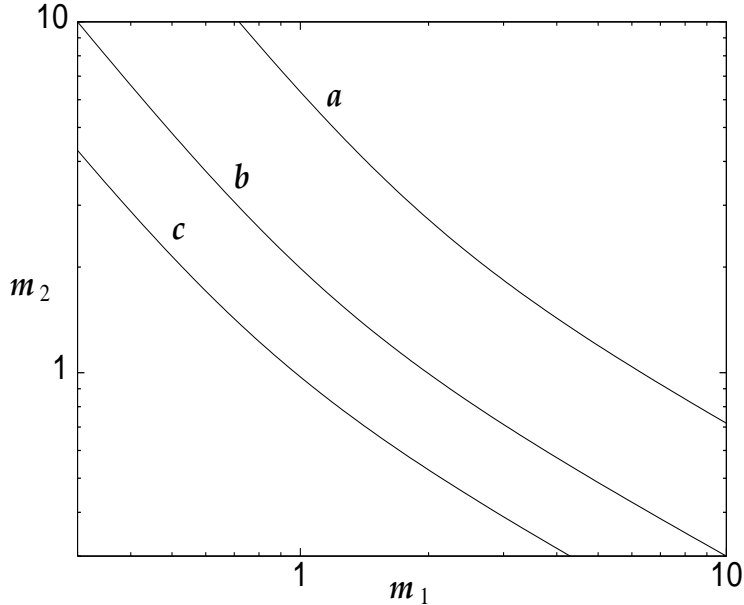,height=12cm,width=10cm,rheight=7.8cm,bbllx=-6cm,bblly=-1.4cm,bburx=11.2cm,bbury=23.6cm}
\caption{Contours of constant $M_c$ for the same hollow sphere 
as in fig. \protect\ref{9}, observing the chirp at the second resonance 
frequency $f_2 = 753$ Hz, with SNR=10. The chirp masses and the maximum 
distances are: $a: M_c = 2.0$, $r=131$ Mpc, $b: M_c = 1.2$, $r=86$ Mpc, 
$c: M_c = 0.9$, $r= 65$ Mpc. If the double passage 
technique is applied, the delay times between the excitation of the first and 
the second mode by the chirps of the given mass are: $a:160$ ms, $b:373$ ms, 
$c:648$ ms.}
\label{10}
\end{figure}

Another consequence of the multimode and multifrequency nature of a
spherically symmetric detector is the possibility to determine the orbit
orientation by the measurement of the relative proportion of the two
polarisation amplitudes, and thereby the distance to the source and the
intrinsic GW amplitudes~\cite{sv}. See figs. \ref{9} and \ref{10} for a
specific example referring to optimally oriented circular orbits.

Because of the Newtonian approximation, eqs. (\ref{ev7}) and (\ref{viv1})
become inaccurate near coalescence. In analogy with previous 
analyses~\cite{bd,sv}, we limit our considerations to the frequency at
which there are still five cycles remaining in the waveform until coalescence.
The highest chirp mass values reported in the figures are determined by the
requirement that the five--cycle--frequency of the source be larger than the
resonant frequencies of the detector.

\subsection{Stochastic background}

In this case $h(t)$ is a random function and we assume that its power spectrum,
indicated by $S_{gw}(f)$, is flat and its energy density per unit logarithmic 
frequency is a fraction $\Omega_{gw}(f)$ of the closure density $\rho_c$ of
the Universe:

\begin{equation}
\frac{d\rho_{gw}}{dlnf} = \Omega_{gw}\rho_c
\end{equation}

$S_{gw}(f)$ is given by

\begin{equation}
S_{gw}(f) = \frac{2G}{\pi}f^{-3}\Omega_{gw}(f)\rho_{c}
\end{equation}

The measured noise spectrum $S_{h}(f)$ of a single resonant-mass
detector automatically gives an upper limit to $S_{gw}(f)$ (and
hence to $\Omega_{gw}(f)$).

Two different detectors with overlapping bandwidth $\Delta f$ will respond to
the background in a correlated way. The SNR of a GW background in a
cross correlation experiment between two detectors located near one another
and having a power spectral density of noise $S_{h}^{1}(f)$ and $S_{h}^{2}(f)$ 
is~\cite{pn}:

\begin{equation}
SNR = \left(\frac{S^{2}_{gw}}{S_{h}^{1}S^{2}_{h}}\Delta f\/t_m\right)^{1/4}
\label{ev11}
\end{equation}

where $t_{m}$ is the total measuring time.

Detectors located some distance apart do not correlate quite so well
because GWs coming from within a certain cone about the line joining the
detectors will reach one of them before the other. The fall-off in the 
correlation with separation is a function of the ratio of the wavelength
to the separation, and has been studied for pairs of bars, pairs of
interferometers~\cite{pp,pq} and pair of spherical detectors \cite{vc}.

Assuming two identical large hollow spherical detectors are co-located
for optimum correlation, the background will reach a {\it SNR\/}\,=\,1
if $\Omega_{gw}$ is 

\begin{equation}
\Omega_{gw}\simeq 10^{-9}\,\times\,\left(\frac{f_{n}}{200\ {\rm Hz}}\right)^{3}
\,\left(\frac{\sqrt{S^1_h(f_n)}}{10^{-24}\ {\rm Hz}^{-1/2}}\right)\,
\left(\frac{\sqrt{S^2_h(f_n)}}{10^{-24}\ {\rm Hz}^{-1/2}}\right)\,
\left(\frac{20\ {\rm Hz}}{\Delta f_n}\right)^{1/2}\,
\left(\frac{10^7\ {\rm sec}}{t_m}\right)^{1/2}
\label{ev12}
\end{equation}

where the Hubble constant has been assumed 100 km\,s$^{-1}$.

Hollow spherical detectors can set very interesting limits on the GW
background. In particular, following recent estimations based on cosmological
string models ~\cite{vz}, it emerges that experimental measurements
performed at the level of sensitivity attainable with these
detectors would be true tests of Planck-scale physics. 

Eqs. (\ref{ev11}) and (\ref{ev12}) hold for whichever cross-correlation
experiment between two GW detectors adjacent and aligned for optimum
correlation. An interesting consequence is that the sensitivity
of a hollow sphere-interferometer observatory will be unprecedented.
It can be worth to build a hollow spherical mass detector close to a 
large interferometer, like LIGO or VIRGO, to perform stochastic 
searches~\cite{pr}.

\section{Conclusions}

In this paper we have been mainly concerned with the problem of how an elastic
{\it hollow\/} sphere responds to a GW signal impinging on it. To address this
problem we have developed an analytical procedure to fully sort out the
eigenfrequencies and eigenmodes of that kind of solid, then applied it to 
calculate the GW {\it absorption cross section\/} for arbitrary thicknesses 
and materials of our solid.

When realistic hypotheses are made regarding the size and material of a
possible GW detector of this shape, we have seen that a hollow sphere can be
advantageous in several respects. It has all the features associated with its
symmetry, such as omnidirectionality and capability to determine the source
direction and wave polarisation. Also, its quadrupole frequencies are below
those of an equally massive solid sphere, thus making the low frequency range
accessible to this antenna with good sensitivity. We have investigated the
system response to the classical GW signal sources (bursts, chirps, continuous
and stochastic) for several sizes and materials, and seen that interesting
signal-to-noise ratios are attainable with such a detector. Also, its
bandwidth partly overlaps with that of the projected large interferometers
\cite{vir,lig}, so potentially both kinds of detectors can be operated
simultaneously to make {\it hybrid\/} GW observatories of unprecedented
sensitivity and signal characterisation power.

While it seems possible to cool a 100 ton solid sphere down to 50 mK \cite{tf}, 
the possibility of cooling a large hollow sphere at such low temperatures, as 
well as the fabrication technique and the influence of cosmic rays 
on a low-temperature GW detector of that shape and dimensions,
are currently under investigation.

\acknowledgements{
We are grateful to S. Merkowitz and S. Vitale for helpful discussions, and
also to A. Semeonov from the Kapitza Institute for Physical Problems,
Moscow. One of us (GF) has received financial support from the Dutch NWO,
two of us (JAL and JAO) from the Spanish Ministry of Education through
contract number PB93-1050, and two of us (EC and VF) from the Istituto
Nazionale di Fisica Nucleare.}

\end{document}